\documentstyle[aps,preprint]{revtex}
\newcommand{\be}{\begin{equation}}
\newcommand{\ee}{\end{equation}}
\newcommand{\ba}{\begin{eqnarray}}
\newcommand{\ea}{\end{eqnarray}}

\newcommand{\non}{\nonumber\\}
\newcommand{\eq}[1]{(\ref{#1})}

\newcommand{\cc}{\beta}
\newcommand{\cd}{{\beta_1}}
\newcommand{\dd}{\gamma}

\newcommand{\hil}{{\bf\Omega}}
\newcommand{\hils}{{{\bf\Omega}^s}}

\newcommand{\Int}{{\Upsilon}}

\newcommand{\Hc}{{\bar{H}_{C}}}
\newcommand{\Hcm}{{\bar{H}_{CM}}}
\newcommand{\Hpf}{{\bar{{\cal H}}_{PF}}}

\newcommand{\cD}{{\cal D}}
\newcommand{\hD}{\widehat{D}}

\renewcommand{\d}{{\bf\Xi}}
\renewcommand{\mid}{{\mit\Xi}}

\newcommand{\ex}[1]{{M_{#1}}}
\newcommand{\re}[1]{{M_{#1}}}
\newcommand{\exb}[1]{{\overline{M}_{#1}}}

\newcommand{\sex}[1]{{P_{#1}}}
\newcommand{\sre}[1]{{P_{#1}}}
\newcommand{\sexb}[1]{{\overline{P}_{#1}}}

\newcommand{\del}[1]{{\partial_{#1}}}

\newcommand{\cO}{{\cal P}}
\newcommand{\cI}{{\cal I}}

\newcommand{\bR}{{\bf \mbox{R}}}

\newcommand{\bC}{{\bf \mbox{C}}}

\begin{document}
%%%%%%%%%%%%%%%%%%%%%%%%%%%%%%%%%%%%%%%%%%%%%%%%%%%%%%%%%%%

%%%
\draft
%%%
\preprint{%
\vbox{%
 \hbox{\hspace*{13.7cm}YITP/96-7\hspace*{2cm}}
 \raisebox{1cm}{%
  \vbox{% 
    \hbox{\hspace*{12.1cm}{\tt UT-Komaba/96-4}\hspace*{2cm}}
    \raisebox{1cm}{\hspace*{11cm}{\tt cond-mat/9602105}\hspace*{2cm}}
       }
 \hbox{}
%\hbox{}
}
}
}

%%%%%%%%%%%%%%%%%%%%%%%%%%%%%%%%%%%%%%%%%%%%%%%%%%%%%%%%%%%
\title{%
Integrable $1/r^2$ Spin Chain with 
Reflecting End
}
\author{Takashi Yamamoto\thanks{
{\it E-mail address}
: yam@yukawa.kyoto-u.ac.jp
}
}
\address{%
Yukawa Institute for Theoretical Physics, 
Kyoto University, \\
Kyoto 606, Japan}
\author{Osamu Tsuchiya\thanks{
{\it E-mail address}
: otutiya@hep1.c.u-tokyo.ac.jp}}
\address{%
Department of Pure and Applied Sciences,
University of Tokyo,\\
Komaba, Meguro-ku, Tokyo 153, Japan
}
%\date{\today}
%%%
\maketitle
%%%
\begin{abstract}
A new integrable spin chain of the Haldane-Shastry type
is introduced.
It is interpreted as the inverse-square interacting spin chain 
with a {\it reflecting end}.
The lattice points of this model consist of the square roots
of the zeros 
of the Laguerre polynomial.
Using the ``exchange operator formalism'',
the integrals of motion for the model are explicitly constructed.
\end{abstract}
%%%
%%%

%%%%%%%%%%%%%%%%%%%%%%%%%%%%%%%%%%%%%%%%%%%%%%%%%%%%%%%%%

%%Intro

Studies of the Calogero-Sutherland model \cite{CS},
the Haldane-Shastry spin chain \cite{HS} 
and their variants \cite{revCS}
have provided many new links with other areas 
of physics and mathematics.
In particular, these models provide exactly solvable models in which
the ideas of 
the fractional exclusion statistics can be tested \cite{EX,BW}.

In ref. \cite{exchange}, 
with a view to proving the quantum integrability of 
the Calogero-Sutherland model and, its rational version, 
the Calogero-Moser model
confined in a harmonic potential 
(we call the Calogero model),
Polychronakos had proposed 
the so-called {\it exchange operator formalism}.
His clever formalism could be applicable 
not only to the continuum models but to the lattice models
and has become a standard technique to study 
the integrability and the spectrum of 
the inverse-square interacting systems \cite{MPb,FM,BHV,BV,DKV,UW}.
Within the exchange operator formalism,
all of the inverse-square interacting lattice models
can be related to the appropriate continuum 
inverse-square interacting models
with the {\it internal degrees of freedom (spin)}.
More precisely, the lattice models are obtained by freezing out 
the kinematic degrees of freedom 
in the corresponding continuum models,
and lattice sites lie at the classical static-equilibrium positions 
of the continuum models \cite{zero,zero-gene,zero-L}.
For example \cite{FM,PFP}, the Haldane-Shastry model is related 
to the spin Calogero-Sutherland model \cite{HaHa,MPb,HW}
whose classical equilibrium positions form a regular lattice on the circle.

Polychronakos \cite{PFP} has applied his formalism to constructing
the new lattice model related to 
the spin Calogero model \cite{MPb,BV,DKV,H-W93,VOK}.
We call this model the Polychronakos-Frahm (PF) model \cite{PFF,PFH}. 
The lattice sites of the PF model  
are positioned at the zeros of 
the Hermite polynomial, {\it i.e.}, 
the spins are no more equidistant.
Against this unusual property, the spectra of the PF model 
are equally spaced and therefore simpler 
than those of the Haldane-Shastry model. 
Thus the fractional exclusion statistics 
for the elementary excitations of the PF model 
is more tractable than one of the Haldane-Shastry model \cite{PFH}.

On the other hands, in ref. \cite{S-A94,B-P-S},
an another generalization of the spin chain model,
the Haldane-Shastry model with open boundary conditions
($BC_N$-type Haldane-Shastry model), 
has been introduced. 
This model is related to 
the $BC_N$-type spin Calogero-Sutherland model \cite{O-Pa,Chere}.
It is now well known that such $BC_N$-type models 
can be applicable to analyzing the physics 
with boundaries \cite{K-S94,B-R,Caselle,yky}.
In particular, one of the authors and his collaborators have shown
that the above models
possess the properties of 
the chiral Tomonage-Luttinger liquids \cite{yky}.

The aim of this letter is twofold.
The first is to prove the integrability of 
the $B_N$-type spin Calogero model \cite{yam} within 
the exchange operator formalism.
The second is 
to construct 
the new integrable lattice model related to the 
$B_N$-type spin Calogero model.
This lattice model is thought of as 
the ``intersection'' of 
the PF model and the $BC_N$-type Haldane-Shastry model.

Before turning to the explicit calculation,
we shall briefly mention this new integrable spin chain.
The Hamiltonian 
is given by,
\be
\label{b-pfm}
{{\cal H}_{PF}}
=
\sum_{1\leq j\ne k\leq N}
\left[
\frac{1}{(x_j-x_k)^2}\sex{jk}
+
\frac{1}{(x_j+x_k)^2}\sexb{jk}
\right]
+
\dd\sum_{j=1}^N \frac{1}{x_j^2}\sre{j},
\ee
where $N$ is the number of sites
and $\gamma\in \bR$ is a parameter.
In the above Hamiltonian we have introduced the 
$B_N$-type spin exchange operators
for the $\nu$-component spin variables \cite{B-P-S,yam};
the operator $\sex{jk}$ exchanges the spins 
at the sites $j$ and $k$, the operator $\sre{j}$ 
is defined by the condition $\sre{j}^2=1$ and thus 
is regarded as a reflection operator of the spin at the site $j$,
and finally the operator $\sexb{jk}$ is defined by 
$\sexb{jk}=\sre{j}\sre{k}\sex{jk}$.
Also it will be shown that, from 
the integrability condition of the model,
lattice points $x_j$'s lie at the square roots of the zeros of 
the Laguerre polynomial $L_N^{(|\gamma|-1)}(y)$
(see, for the notation, ref. \cite{ortho}). 
It is well known that the Laguerre polynomial $L_N^{(\alpha)}(y)$
with $\alpha> -1$ (resp. $=-1$) has $N$ distinct roots,
$0<y_1<y_2<\cdots<y_N$ (resp. $0=y_1<y_2<\cdots<y_N$) \cite{ortho}. 
Therefore the lattice of the model is well defined 
and does not contain negative sites. For example, in the case 
$N=4, \dd=2$, the model has the lattice (0.86, 1.60, 2.39, 3.31).

There are several points which should be noticed in \eq{b-pfm}.
Clearly, the Hamiltonian \eq{b-pfm} is not translationally invariant
because the lattice is not uniform. 
Even if we suppose that the lattice is uniform,
the terms $\sexb{jk}/(x_j+x_k)^2$ and $\sre{j}/x_j^2$ in \eq{b-pfm}
break the translational invariance.
The term $\sexb{jk}/(x_j+x_k)^2$ represents the interaction
between the $j$-th. spin and the ``mirror-image'' of 
the $k$-th. spin.
With an appropriate choice of 
the representation of the operator $\sre{j}$,
the last term of \eq{b-pfm} can be regarded as magnetic fields
whose magnitudes are proportional to the inverse-square 
of the positions of the sites. From these observations,
the origin $x=0$ can be regarded as 
a reflecting end of the system.
Then we call the model with Hamiltonian \eq{b-pfm} 
the PF model with reflecting end or 
the $B_N$-PF model
(if $\dd=0$, we call the $D_N$-PF model).

Consider now the integrability of the $B_N$-type spin Calogero model.
We first recall the $B_N$-type spin Calogero model.
The Hamiltonians of the 
$B_N$-type spin Calogero-Moser model and the $B_N$-type spin Calogero model
are respectively given by \cite{yam},
\ba
\label{cmm}
\Hcm
&=&
\sum_{j=1}^N
\left[
-\del{j}^2
+
\frac{1}{x_j^2}\cd(\cd-\re{j})
\right]
\non
&+&\sum_{1\leq j\ne k\leq N}
\left[
\frac{1}{(x_j-x_k)^2}\cc(\cc-\ex{jk})
+
\frac{1}{(x_j+x_k)^2}\cc(\cc-\exb{jk})
\right],
\\
\label{cm}
\Hc&=&\Hcm+\omega^2\sum_{j=1}^Nx_j^2,
\ea
where $\cc, \cd\in\bR$ and 
$\omega\in\bR_{\geq0}$ are coupling constants, 
and $\del{j}=\frac{\partial}{\partial x_j}$.
In \eq{cmm}, we have already introduced 
the operators $\re{j},\ex{jk}$
and $\exb{jk}(=\re{j}\re{k}\ex{jk})$ which are called the $B_N$-type 
(coordinate) exchange operators,
and are defined by the action on the coordinates $x_j$;
\be
\ex{jk}x_j=x_k\ex{jk},\  \re{j}x_j=-x_j\re{j}.
\ee
It is easy to see that these operators satisfy the following relations;
\ba
&&
\re{j}^2=\ex{jk}^2=\exb{jk}^2=1,
\\
&& 
\ex{jk}=\ex{kj},\ \exb{jk}=\exb{kj},
\\
&&
\re{j}\re{k}=\re{k}\re{j},
\\
&&
\ex{jk}\re{j}=\re{k}\ex{jk},\ 
\exb{jk}\re{j}=\re{k}\exb{jk}=\ex{jk}\re{k},
\\
&&
\ex{jk}\ex{kl}=\ex{kl}\ex{jl}=\ex{jl}\ex{jk},
\\
&&
\ex{jk}\exb{kl}=\exb{jl}\ex{jk}=\exb{kl}\exb{jl}.
\ea
Remark that the $B_N$-type spin exchange operators 
$\sre{j}, \sex{jk}$ and $\sexb{jk}$ 
also satisfy the above relations \cite{yam}.

The Hamiltonians \eq{cmm} and \eq{cm} does not 
contain the terms related directly to the spin.
The spin degrees of freedom are introduced as follows.
Let
$\hils=C^\infty(\bC^N)\otimes V$ 
where $V$ denotes the space of spins,
for example, $(\bC^\nu)^{\otimes N}$.
Then operators $\ex{jk}, \re{j}, \sex{jk}$ and $\sre{j}$ 
naturally act on this space,
and clearly $\ex{jk}$ and $\re{j}$ commute with
$\sex{jk}$ and $ \sre{j}$. 
Next we introduce a projection $\pi$ 
which  respectively replaces every occurrence of 
$\ex{jk}$ and $\re{j}$ by $\sex{jk}$ and $\sre{j}$ 
after $\ex{jk}$ and $\re{j}$ have been moved to 
the right of the expression.
Consider the $B_N$-type ``bosonic'' subspace
\ba
\label{hils}
\widetilde{\hil}^s
&=&\{f\in\hils\  | \ (\ex{jk}-\sex{jk})f=0,\ (\re{j}-\sre{j})f=0\}.
\ea
For any operator $\bar{{\cal O}}$, 
the projection $\pi$ leads to a unique operator ${\cal O}$ 
which satisfies 
$\bar{{\cal O}}\widetilde{\hil}^s={\cal O}\widetilde{\hil}^s$
and does not contain the coordinate exchange operators.
The Hamiltonians with the spin degrees of freedom 
are thus given by
the operators $\pi(\Hcm)$ and $\pi(\Hc)$.
Also, the spinless, {\it i.e.}, the one-component case can be 
considered by putting $\sex{jk}=1, \sre{j}=1$.
In this case,
the conditions in \eq{hils} are nothing but the conditions for
the $B_N$-invariance of the wavefunctions.

First of all, we introduce the operators $\cD_j$ for later use;
\be
\cD_j=
\sum_{k\ne j}
\left[
\frac{1}{x_j-x_k}\ex{jk}
+
\frac{1}{x_j+x_k}\exb{jk}
\right]
+
\frac{\cd}{\cc}
\frac{1}{x_j}\re{j}.
\ee
It is easy to show that
\ba
&&
\re{j}\cD_j=-\cD_j\re{j},\
\ex{jk}\cD_j=\cD_k\ex{jk},\ 
\\
&& 
[\cD_j,\cD_k]=0,
\\
&& 
[\cD_j,x_k]
=
\delta_{jk}
\left(
-\sum_{l\ne j}(\ex{jl}+\exb{jl})
-2\frac{\cd}{\cc}\re{j}
\right)
+
(1-\delta_{jk})(\ex{jk}-\exb{jk}).
\ea
Next, we define the $B_N$-type 
Dunkl operators \cite{Dunkl,Chere,Kuz} $D_j$ by
\be
D_j=\del{j}-\cc \cD_j.
\ee
Using 
$[\del{j},x_j]=\delta_{jk}$,
$\re{j}\del{j}=-\del{j}\re{j},\
\ex{jk}\del{j}=\del{k}\ex{jk}$, etc.,
we can show that the $B_N$-type Dunkl operators $D_j$ 
together with the coordinates $x_j$ 
satisfy the following relations,
\ba
&&%\mbox{(i)}\ 
\re{j}D_j=-D_j\re{j},\
\ex{jk}D_j=D_k\ex{jk},\ 
\\
&&%\mbox{(ii)}\ 
[D_j,D_k]=0,\ [x_j,x_k]=0,
\\
&&%\mbox{(iii)}\ 
[D_j,x_k]
=
\delta_{jk}
\left(
1
+
\cc\sum_{l\ne j}(\ex{jl}+\exb{jl})
+
2\cd\re{j}
\right)
-
(1-\delta_{jk})\cc(\ex{jk}-\exb{jk}).
\ea
Finally we introduce another type of the $B_N$-type Dunkl operators,
\be
D_j^\pm=D_j\mp\omega x_j
\ee
which satisfy the similar relations among $D_j$'s and $x_j$'s;
\ba
&&%\mbox{(i)}\ 
\re{j}D_j^\pm=-D_j^\pm\re{j},\
\ex{jk}D_j^\pm=D_k^\pm\ex{jk},\ 
\\
&&%\mbox{(ii)}\ 
[D_j^\pm,D_k^\pm]=0,
\\
&&%\mbox{(iii)}\ 
[D_j^+,D_k^-]
=
2\omega[D_j,x_k].
\ea
In fact, if we redefine $D_j^\pm$ by $D_j^\pm/\sqrt{2\omega}$,
then $\{D_j,x_j\}$ and $\{D_j^+,D_j^-\}$ have 
the same algebraic structure.

Remark that we can lead to the similar results
starting with the gauge transformed versions
of the $D_j$ and $D_j^\pm$;
\ba
&&\hD_j
=
\Delta(x)^{-1}D_j\Delta(x)
=
D_j
+
\cc\sum_{k\ne j}
\left[
\frac{1}{x_j-x_k}+\frac{1}{x_j+x_k}
\right]
+
\cd\frac{1}{x_j},
\\
&&\hD_j^\pm
=
\widetilde{\Delta}(x)^{-1}
D_j^\pm
\widetilde{\Delta}(x)
=
\hD_j
-
(\omega\pm\omega)x_j,
\ea
where 
$\Delta(x)
=\prod_{j<k}(x_j^2-x_k^2)^\cc\prod_lx_l^{\cd}$
and 
$\widetilde{\Delta}(x)
=\Delta(x)\exp({-\frac{\omega}{2}\sum_j x_j^2})$.

%involutiveness 

As the ordinary case \cite{MPb,PFP},
the integrals of motion for the $B_N$-type (spin) Calogero-Moser model 
and the $B_N$-type (spin) Calogero model 
can be constructed by using the Dunkl operators 
$D_j$ and $D_j^\pm$, respectively.
Moreover, under an appropriate transformation of the coordinates,
the integrals of motion for 
the $B_N$-type (spin) Calogero-Sutherland model
are related to the operators $x_jD_j$.
Then we shall unify the construction of 
these integrals of motion following ref. \cite{BHW}\footnote{%
Precisely speaking, this treatment is not convenient to the case 
of the $B_N$-type (spin) Calogero-Moser model,
because the involutiveness of integrals is clear from its definition.}.
For this purpose, we introduce the operators, 
\be
\d_{j}=(pD_j+q x_j)(p'D_j+q' x_j),
\ee
where $p, p', q, q'\in \bC$. 
They satisfy the relations 
\ba
&&%\mbox{(i)}\ 
\re{j}\d_j=\d_j\re{j},\
\ex{jk}\d_j=\d_k\ex{jk},\ 
\\
&&%\mbox{(ii)}\ 
[\d_j,\d_k]
=
(pq'-p'q)\cc(\d_j-\d_k)(\ex{jk}+\exb{jk}).
\ea
{}From the above formulae we can show 
the key formula,
\be
\label{bf}
[\d_j^n,\d_k^m]
=
(pq'-p'q)\cc
\sum_{a=1}^m \d_k^{m-a}(\d_j^n-\d_k^n)\d_j^{a-1}(\ex{jk}+\exb{jk}).
\ee
Let us consider the quantities
\be
\Int_n=\sum_{j=1}^N \d_j^n.
\ee	
Then the involutiveness of $\Int_n$'s 
is clear if $pq'-p'q=0$.
On the other hand, in general,
using the formula \eq{bf} and then explicitly antisymmetrizing 
in the index,
we can prove the involutiveness of $\Int_n$'s as follows;
\ba
&&[\Int_n,\Int_m]
\non
&=&
(pq'-p'q)\cc\sum_{j,k=1}^N
\sum_{a=1}^m
[
\d_k^{m-a}(\ex{jk}+\exb{jk})\d_k^{n+a-1}
-
\d_k^{n+m-a}(\ex{jk}+\exb{jk})\d_k^{a-1}
]
\non
&=&
(pq'-p'q)\cc\sum_{j,k=1}^N
\left(
\sum_{a=1}^m-\sum_{a=n+1}^{n+m}
\right)
\d_k^{m-a}(\ex{jk}+\exb{jk})\d_k^{n+a-1}
\non
&=&
\frac{(pq'-p'q)\cc}{2}\sum_{j,k=1}^N
\left(
\sum_{a=1}^m-\sum_{a=n+1}^{n+m}
-\sum_{a=1}^n+\sum_{a=m+1}^{n+m}
\right)
\d_k^{m-a}(\ex{jk}+\exb{jk})\d_k^{n+a-1}=0.
\nonumber
\ea
Moreover, from the $B_N$-symmetry of $\Int_n$, {\it i.e.},
$[\ex{jk}, \Int_n]=[\re{j},\Int_n]=0$,
the projections $\pi(\Int_n)$ are also 
involutive.

Specializing the parameters $p, p', q$ and $q'$,
we define the two sets of the involutive operators 
$\{I_n^{CM}\}$ and $\{I_n^{C}\}$ 
corresponding to 
the $B_N$-type spin Calogero-Moser model 
and the $B_N$-type spin Calogero model, respectively;
\ba
&&I_n^{CM}=\Int_n|_{p=p'=1\atop q=q'=0}
=
\sum_{j=1}^N (D_j)^{2n},
%\ \ (pq'-p'q=0)
\\
&&I_n^{C}=\Int_n|_{p=p'=1\atop -q=q'=\omega}
=
\sum_{j=1}^N (D_j^+D_j^-)^n.
%\ \ (pq'-p'q=2\omega)
\ea
Note that, 
in contrast to the ordinary (spin) Calogero-Moser model,
the integrals $I_n^{CM}$ depend only on $D_j^2$.
This fact reflects the absence of the 
translational invariance in the Hamiltonian \eq{cmm}.
Note also 
that $I_n^{CS}=\Int_{n}|_{p=0,p'=1\atop q=-1,q'=0}$  %\ (pq'-p'q=1)
are related to the $BC_N$-type spin Calogero-Sutherland model.

The Hamiltonian $\Hc$ (resp. $\Hcm$)
is expressed by the operator $I_1^{CM}$ (resp. $I_1^{C}$);
\ba
&&\Hcm=-I_1^{CM},
\\
&&\Hc=-I_1^C+{\cal E}_N^{(0)},
\ea
where 
${\cal E}_N^{(0)}
=\omega[N+2\cc\sum_{j<k}(\ex{jk}+\exb{jk})+2\cd\sum_j\re{j}]$.
It remains to show that $I_n^{CM}$'s (resp. $I_n^{C}$ 's)
commute with $\Hcm$ (resp. $\Hc$).
These can be checked by using the formulae;
\ba
\label{d-Hcm}
&&[\Hcm,D_j]=0,
\\
\label{sga-c}
&&[\Hc,D_j^\pm]=\pm 2\omega D_j^\pm.
\ea							
Hence 
the $B_N$-type spin Calogero-Moser model 
and the $B_N$-type spin Calogero model are integrable.
As mentioned, using the projection $\pi$,
we can obtain the corresponding integrals of motion
which depend on the spin variables.

%%%%%%%%%%%%%%%%%%%%%%%%%%%%%%%%%%%%%%%%%%%%

Let us now turn to the lattice model
related to the $B_N$-type spin Calogero model.
We apply the standard technique due to Polychronakos \cite{PFP}
(see also \cite{SS93,kaku}).
That is, we consider the strong coupling limit 
$\cc \rightarrow \infty$ in the Hamiltonian \eq{cm}.
Since the repulsion between particles and also 
between particles and mirror-image particles
become dominant in the strong coupling limit,
particles are enforced to localize with the positions $x_j$
which are taken to minimize the potential,
\be
V(x)
=
\cc^2\tilde{\omega}^2\sum_{j=1}^Nx_j^2
+
\cc^2\sum_{1\leq j\ne k\leq N}
\left[
\frac{1}{(x_j-x_k)^2}
+
\frac{1}{(x_j+x_k)^2}
\right]
+
\cc^2\dd^2\sum_{j=1}^N
\frac{1}{x_j^2}.
\ee
Here we rescaled the the coupling constant $\omega$ of the harmonic potential 
in order for the system to have a nontrivial limit.
Also we rescaled $\cd=\cc\dd$. 
Note that 
$\tilde{\omega}$ can be absorbed into the definition of $x_j$'s.
Then we put $\tilde{\omega}=1$. From $\del{j}V(x)=0$,
we can obtain that such $x_j$'s satisfy the condition
\be
\label{la-zero}
2\sum_{k\ne j}
\left[
\frac{1}{(x_j-x_k)^3}+\frac{1}{(x_j+x_k)^3}
\right]
+
\dd^2\frac{1}{x_j^3}
=
x_j.
\ee
The above formula is equivalent to the condition
that $y_j=x_j^2$ are zeros of 
the Laguerre polynomial $L_N^{(|\dd|-1)}(y)$  \cite{zero-L}.

In the strong coupling limit $\cc\rightarrow\infty$,
the elastic modes decouple from the internal degrees of freedom
(the latter constitute the desired spin chain model);
\be
\Hc\longrightarrow H_{ela}-\cc\Hpf.
\ee
Here
$H_{ela}$ represents the Hamiltonian for the elastic degrees 
of freedom
and 
$\Hpf$ is the Hamiltonian 
which is obtained by 
replacing 
$\sex{jk}$ and $\sre{j}$ 
respectively with $\ex{jk}$ and $\re{j}$
in \eq{b-pfm}, {\it i.e.}, ${{\cal H}_{PF}}=\pi(\Hpf)$.

Let us define the operators
\ba
&&
\cD_j^\pm=\cD_j\pm x_j,
\\
&&
\mid_j
=\cD_j^+\cD_j^-
=\cD_j^2-x_j^2-\sum_{k\ne j}(\ex{jk}+\exb{jk})-\dd\re{j}.
\ea
The operators $\cD_j^\pm$ 
can be thought of as 
the large-$\cc$ limit of the operators $D_j^\pm$.
Thus we expect that 
the operators 
$
\cI_n^{PF}=\sum_{j=1}^N\mid_j^n
$
are the integrals of motion for the $B_N$-PF model.
We can show the involutiveness of the operators $\cI_n^{PF}$ 
along the same argument as 
those for the $B_N$-type spin Calogero model.
The remaining task is to show the commutativity of 
$\cI_n^{PF}$ with $\Hpf$. 
Clearly,
it suffices to show $[\Hpf,\mid_j]=0$.
This can be proved as follows.
We recall the relation \eq{d-Hcm},
\ba
\label{qq}
[\Hcm,D_j]=0
\Longleftrightarrow
[-\sum_l\del{l}^2-\cc\Hpf+\cc^2\cO,\del{j}-\cc\cD_j]=0,
\ea
where  
\be
\cO=\sum_{1\leq j\ne k\leq N}
\left[
\frac{1}{(x_j-x_k)^2}+\frac{1}{(x_j+x_k)^2}
\right]
+
\dd^2\sum_{j=1}^N\frac{1}{x_j^2}.
\ee
Let us consider the expansion of the relation \eq{qq} 
in the power of $\cc$.
Since this relation holds for all $\cc$,
each term must separately vanish.
Thus the term of the order $\cc^2$ gives,
\be
\label{Hpf-d}
[\Hpf,\cD_j]
=[\del{j},\cO]
=
-4\sum_{k\ne j}
\left[
\frac{1}{(x_j-x_k)^3}+\frac{1}{(x_j+x_k)^3}
\right]
-
2\dd^2\frac{1}{x_j^3}.
\ee
Also the direct calculation show that 
\be
\label{Hpf-x}
[\Hpf,x_j]
=-2\cD_j.
\ee
Using the above two formulae \eq{Hpf-d}, \eq{Hpf-x}
and 
the properties $[\Hpf,\ex{jk}]=[\Hpf,\re{j}]=0$,
we obtain,
\be
[\Hpf,\mid_j]
=
([\Hpf,\cD_j]+2x_j)\cD_j+\cD_j([\Hpf,\cD_j]+2x_j).
\ee
If $x_j$'s are chosen to take values in the set of square 
roots of the zeros of the Laguerre polynomial 
$L_N^{(|\dd|-1)}(y)$, 
then we have $[\Hpf,\cD_j]+2x_j=0 \ 
(\Leftrightarrow\eq{la-zero})$, 
hence $[\Hpf,\mid_j]=0$.

Therefore 
we proved the integrability of the $B_N$-PF model
and 
obtained the integrals of motion $\pi(\cI_n^{PF})$ 
for this model.
For example, $\pi(\cI_1^{PF})$ is given by,
\ba
\pi(\cI_1^{PF})
=
-E_{N}
-
\left[
\sum_{1\leq j\ne k\leq N}(\sex{jk}+\sexb{jk})
+2\dd\sum_{j=1}^N\sre{j}
\right],
\ea
where
\ba
E_{N}
=
\sum_{j=1}^Nx_j^2
+\sum_{1\leq j\ne k\leq N}
\left[\frac{1}{(x_j-x_k)^2}+\frac{1}{(x_j+x_k)^2}\right]
+\dd^2\sum_{j=1}^N\frac{1}{x_j^2}.
\ea

%%%%%%%%%%%%%%%%%%%%%%%%%%%%%%%%%%%%%%%%%%%%%%%%%

Finally, 
we would like to make some comments on algebraic 
interpretations of the presented results. 
Our construction naturally leads to the algebra of 
integrals of motion.
For example, 
the Virasoro-like structure is given by
%Virasoro algebra without the central extension
\ba
&&
[J_n,J_m]=0,
\\
&&
[L_n,J_m]=-mJ_{n+m},
\\
&&
[L_n,L_m]=(n-m)L_{n+m},
\ea
where 
\ba
\label{current}
J_n&=&I_n^{CM},\ (\mbox{or}\ \ I_n^{C}/(2\omega)^n),
\\
\label{virasoro}
L_n&=&\frac{1}{2}\sum_{j=1}^Nx_jD_j^{2n+1},\ 
\left(
\mbox{or}\ \ 
\frac{1}{2}\sum_{j=1}^ND_j^-(D_j^+)^{2n+1}/(2\omega)^{n+1}\right).
\ea
For the proof, we used the formula,
\ba
&&[D_j^n,x_k]
\non
&=&
\delta_{jk}
\left[
nD_j^{n-1}
+
\cc\sum_{l\ne j}
\left(
P^-(D_j,D_l)\ex{jl}+P^+(D_j,D_l)\exb{jl}
\right)
+
(1-(-1)^n)\cd D_j^{n-1}\re{j}
\right]
\non
&- &
(1-\delta_{jk})\cc
\left[
(P^-(D_j,D_k)\ex{jk}-P^+(D_j,D_k)\exb{jk}
\right],
\ea
where the polynomials $P^\pm(X,Y)$ are defined by
$P^\pm(X,Y)=(X^n\pm Y^n)/(X\pm Y)$.
Notice that in \eq{virasoro}
the total degree of the operator is always even 
as the polynomial of $x_j$ and $D_j$ 
(or $D_j^-$ and $D_j^+$),
this fact reflects the $B_N$-symmetry.
We can also construct the algebra 
of integrals of motion
related to the $W_\infty$ algebra.

Another important futures are relations
to the spectrum generating algebras and the Yangian symmetries.
One of the authors has shown that 
the spectrum of the $B_N$-type spin Calogero model
is equally spaced \cite{yam}.
It is easy to see that the same is true for the $B_N$-PF model.
This is caused by the existence of the 
spectrum generating algebras \eq{sga-c}
and 
\be
\label{sga-pf}
[\Hpf, \cD_j^\pm]=\mp2\cD_j^\pm. 
\ee
Moreover the numerical studies show 
that the $B_N$-PF model possesses the ``super-multiplet'' structure.
The algebra underlying this structure is 
Yangian \cite{HHTBP,BGHP,PFH,BHW}. 
The Yangian symmetries of the $B_N$-type spin Calogero model
and the $B_N$-PF model are easily see from 
the transfer matrices of these systems which can be constructed
by the Dunkl operators $D_j^\pm$ and $\cD_j^\pm$.
The details will be appeared in \cite{tb}.

{\bf Acknowledgments}
\vskip 2mm
\noindent

We would like to thank 
H.~Awata, N.~Kawakami, Y.~Matsuo, S.~Odake 
and S.-K. Yang
for discussions. 		
T.Y. was supported by 
the COE (Center of Excellence) researchers program
of the Ministry of Education, Science and Culture, Japan.

%%%%%%%%%%%%%%%%%%%%%%%%%%%%%%%%%%%%%%

%%%%%%%%%%%%%%%%%%%%%%%%%%%%%%%%%%%%%%%%%%%%%%%%%%%%%%%%
\end{document}